\documentclass[a4paper]{jpconf}
\usepackage{graphicx}
\usepackage{xcolor}
\usepackage{amsmath}
\usepackage{amsfonts}
\usepackage[colorlinks=true,allcolors=magenta]{hyperref}

\begin{document}
\title{Cosmological radiation density with non-standard neutrino-electron interactions}

\author{Pablo Mart\'inez-Mirav\'e}

\address{Departament de F\'isica  Te\'orica,  Universitat  de  Val\`{e}ncia, and Instituto de F\'{i}sica Corpuscular, CSIC-Universitat de Val\`{e}ncia, 46980 Paterna, Spain}

\ead{pamarmi@ific.uv.es}

\begin{abstract}
Non-standard interactions (NSI) between neutrinos and electrons can significantly modify the decoupling of neutrinos from the plasma. These interactions have two effects on the overall picture: (i) they alter neutrino oscillations though matter effects and (ii) they modify the scattering and annihilation processes involving neutrinos and electrons and positrons. We study the role of non-universal and flavour-changing NSI in the decoupling and how they impact the determination of the effective number of neutrinos, $N_{\rm eff}$. We examine the degeneracies between NSI parameters and we compare the expected sensitivity from future cosmological surveys with the current limits from terrestrial experiments. We outline the complementarity between both approaches.
\end{abstract}

\section{Introduction}

The phenomenon of neutrino oscillations is a clear indication that there is physics beyond the Standard Model (SM) in the neutrino sector. The measurements of flavour oscillations performed using solar, reactor, atmospheric and accelerator neutrinos are well understood in the three neutrino framework with non-zero masses and mixing (see for instance \cite{deSalas:2020pgw}). In many theoretical extensions of the SM, including neutrino mass models, the existence of additional interactions among neutrinos and with other fermions are predicted. There are two main consequences of these non-standard interactions (NSI). In the first place, the propagation of neutrinos in a medium is modified, as it happens due to electroweak interactions. Consequently, NSI give rise to an effective potential that can alter neutrino oscillations. Secondly, these interactions would also give rise to distinguishable signatures in scattering experiments. This scenario has been extensively studied and since there is no evidence of non-zero NSI, multiple constraints have been derived. Non-standard interactions are also expected to play a role in astrophysical processes as well as in cosmology. In these proceedings, we focus on the latter and in particular, on the impact of NSI in the process of neutrino decoupling.

It is well-known that when the temperature fell below the MeV scale, the weak interactions responsible for keeping neutrinos in thermal equilibrium with the plasma stopped being effective. As a consequence, neutrinos decoupled and formed the cosmic neutrino background. This complicated process has been studied with great detail in the framework of the Standard Model, taking into account all neutrino interactions and finite-temperature corrections to Quantum Electrodynamics (QED). The effect of neutrino oscillations has also been included in such calculations \cite{Froustey:2020mcq,Bennett:2020zkv}.

After the epoch of electron-positron annihilation, the radiation density of the Universe can be expressed as
\begin{equation}
    \rho_{\rm rad} = \rho_\gamma \Bigg\{ 1 + \frac{7}{8} \left(\frac{4}{11}\right)^ {4/3} N_{\rm eff}\Bigg\}\, ,
\end{equation}
where we have introduced the effective number of neutrinos, $N_{\rm eff}$, as a parameter that quantifies the ratio between the energy density in form of neutrinos and photons. According to the latest calculations, the value of the effective number of neutrinos is found to be $3.0440$ \cite{Froustey:2020mcq, Bennett:2020zkv}. Nonetheless, deviations from this value are expected in the presence of additional relativistic degrees of freedom or new physics in the neutrino sector. 

From a combination of the measurements of the anisotropies in the cosmic microwave background (CMB) performed by Planck with other cosmological datasets, one determines $N_{\rm eff} = 2.99^{+0.34}_{-0.33}$ at $95\%$ C.L. \cite{Planck:2018vyg}. Such value already constrains significantly the possibility of having new degrees of freedom. In the light of the forecasted sensitivity of forthcoming CMB observatories such as the CMB-S4 project ($\sigma(N_{\rm eff}) = 0.02 - 0.03 $) \cite{CMB-S4:2016ple} and the Simons Observatory ($\sigma(N_{\rm eff})= 0.05 - 0.07$) \cite{SimonsObservatory:2018koc}, we study the impact of NSI in the decoupling of neutrinos and in the determination of $N_{\rm eff}$, extending previous analysis \cite{Mangano:2006ar,deSalas:2016ztq}.

\section{Non-standard interactions and current bounds}
In this work, we only consider neutral current non-standard interactions (NC-NSI) between neutrinos and electrons. The  reason for this choice is twofold. Firstly, constraints on NC-NSI are considerably weaker than the charged current ones. Secondly, we are interested in the the decoupling of relic neutrinos, fo which only electron-neutrino NSI are estimated to be relevant. The effective Lagrangian parametrising the interactions between neutrinos and electrons in this scenario is 
\begin{flalign}
    \mathcal{L}_{\nu e} = -2\sqrt{2}G_F \left[\left(\bar{\nu}_e \gamma ^\mu P_L e\right)\left(\bar{e}\gamma_\mu P_L\nu_e\right) + \sum _{\alpha, X} g_X\left(\bar{\nu}_{\alpha}\gamma^\mu P_L\nu_\alpha\right)\left(\bar{e}\gamma_\mu P_Xe\right) \right.  \nonumber \\ + \left. \sum_{\alpha, \beta, X}\varepsilon_{\alpha \beta} ^ X \left(\bar{\nu}_{\alpha} \gamma^\mu P_L \nu_\beta \right)\left(\bar{e}\gamma_\mu P_L e \right)\right] \, ,
\end{flalign}
where the first two terms account for charged current and neutral current interactions in the SM, respectively, and parametrises NC-NSI. In the effective Lagrangian, $G_F$ is the Fermi constant, greek indices indicate the three different neutrino flavours and $X = \lbrace L\, , R \rbrace$, so that $P_X$ are the left-handed and right-handed chiral projectors.  The strength of the neutral current interactions in the Standard Model depends on the parameters $g_L = \sin^2\theta_W -1/2$ and $g_R = \sin^2\theta_W$, where $\theta_W$ is the weak angle. In the case of NSI, their strength is determined by the dimensionless coefficients $\varepsilon_{\alpha \beta}^X$.

Non-standard interactions can be classified in two categories. We refer to $\varepsilon_{\alpha \beta}^ X$, with $\beta \neq \alpha$, as flavour-changing NSI, since for non-zero values of these parameters, lepton flavour symmetry is broken. The coefficients $\varepsilon_{\alpha\alpha}^ X$ account for non-universal NSI, since in case the difference between two of them is different from zero, $\varepsilon_{\alpha\alpha}^X - \varepsilon_{\beta\beta}^X \neq 0$,  lepton flavour universality does not hold.

Current limits on non-standard interactions come from a variety of terrestrial experiments. They rely on the fact that NSI modify the cross-section measured at scattering experiments. They also affect neutrino propagation (through matter effects) so that they can be probed in oscillation experiments.  For a recent review on the existing bounds, see \cite{Farzan:2017xzy}.

\section{Non-standard interactions and Neff - what to expect}
Non-standard interactions affect the thermalisation of neutrinos in two different ways. In the first place, the modify neutrino oscillations through matter effects. Since the change on $N_{\rm eff}$ arising from neutrino oscillations is known to be of the order of 0.001 (see \cite{Bennett:2020zkv}), we expect that the modification introduced in flavour oscillations by NSI will be negligible. Nonetheless, they are included in our calculations and we will discuss them in the following sections. The second modification in the picture of neutrino decoupling is due to the changes in the collision integrals encoding the interactions involving neutrinos and electrons/positrons. The scattering and annihilation processes involved are proportional to the coefficients $g_L^ 2$, $g_R^ 2$ and the combination $g_L g_R$. The presence of NSI introduces the following shifts in the coefficients:
\begin{flalign}
g_L^2 \longrightarrow \left(g_L + \varepsilon_{\alpha \alpha}^ L\right) ^ 2 + \sum_{\beta\neq \alpha}|\varepsilon_{\alpha \beta}^L|^2
\label{eqn:shifted1}
\end{flalign}
\begin{flalign}
g_R^2 \longrightarrow \left(g_R + \varepsilon_{\alpha \alpha}^ R\right) ^ 2 + \sum_{\beta\neq \alpha}|\varepsilon_{\alpha \beta}^R|^2 
\label{eqn:shifted2}
\end{flalign}
\begin{flalign}
g_L g_R \longrightarrow \left(g_L + \varepsilon_{\alpha \alpha}^ L\right)\left(g_R + \varepsilon_{\alpha \alpha}^ R\right) + \sum_{\beta \neq\alpha}|\varepsilon_{\alpha\beta}^ L||\varepsilon_{\alpha\beta}^ R|.
\label{eqn:shifted3}
\end{flalign}

We use the publicly available code $\texttt{ForEPiaNO}$ \footnote{\url{ https://bitbucket.org/ahep_cosmo/fortepiano_public}} \cite{Gariazzo:2019gyi} to compute the evolution of the density matrix $\varrho$ according to
\begin{equation}
    \frac{d}{dx}\varrho(y) = \sqrt{\frac{3M_{\rm Pl}^2 }{8\pi \rho}} \Bigg\{ \frac{-ix^3}{m_e^3} \left[\frac{\mathbb{M}_F}{2y} - \frac{2\sqrt{2}G_Fym_e^2}{x^6} \left(\frac{\mathbb{E}_l + \mathbb{P}_l}{m_W^ 2} + \frac{4 \mathbb{E_\nu}}{3m_Z^ 2}\right), \varrho\right] + \frac{m_e^3}{x^ 4}\mathcal{I}(\varrho)\Bigg\}
\end{equation}
where we have introduced the comoving variables $x = m_e a$, $y = p a$ and $z = T_\gamma a$, which depend on the electron mass $m_e$, the neutrino momentum $p$, the photon temperature $T_\gamma$, and the scale factor $a$. In the evolution equation above, $M_{\rm Pl}$ denotes the Planck mass, $m_Z$ and $m_W$ are the masses of the Z and W bosons respectively and $\rho$ is the total energy density of the universe. In addition, $\mathbb{M}_F$ is the neutrino mass matrix in the flavour basis, $\mathbb{E}_{l}$, $\mathbb{P}_l$ and $\mathbb{E}_\nu$ are the energy density and pressure of charged leptons and neutrinos. Finally, $\mathcal{I}(\varrho)$ accounts for the collision integrals. For a more detailed discussion of the formalism and the procedure employed to compute the evolution of the density matrix, we refer the reader to Refs. \cite{Bennett:2020zkv, Gariazzo:2019gyi}. In our calculations we take the oscillation parameters from \cite{deSalas:2020pgw} and adopt the same numerical settings described in \cite{deSalas:2021aeh}, where it was shown that the contributions from muons can be safely ignored.

\begin{figure}
    \centering
    \includegraphics[width= 0.485\textwidth]{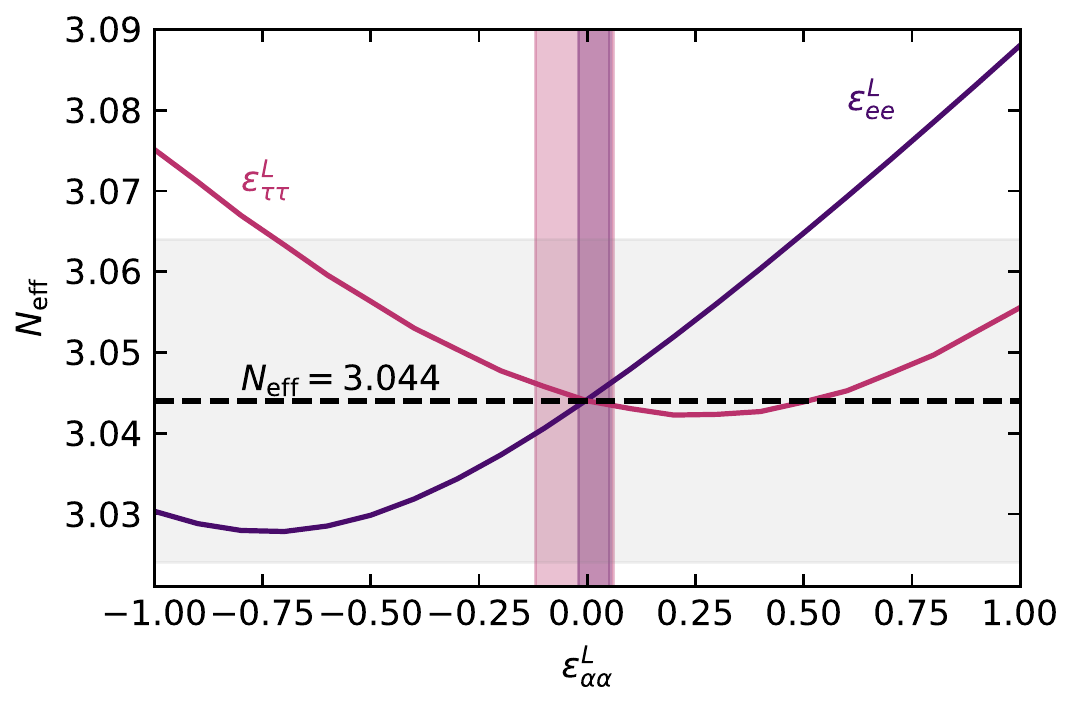}
    \includegraphics[width= 0.485\textwidth]{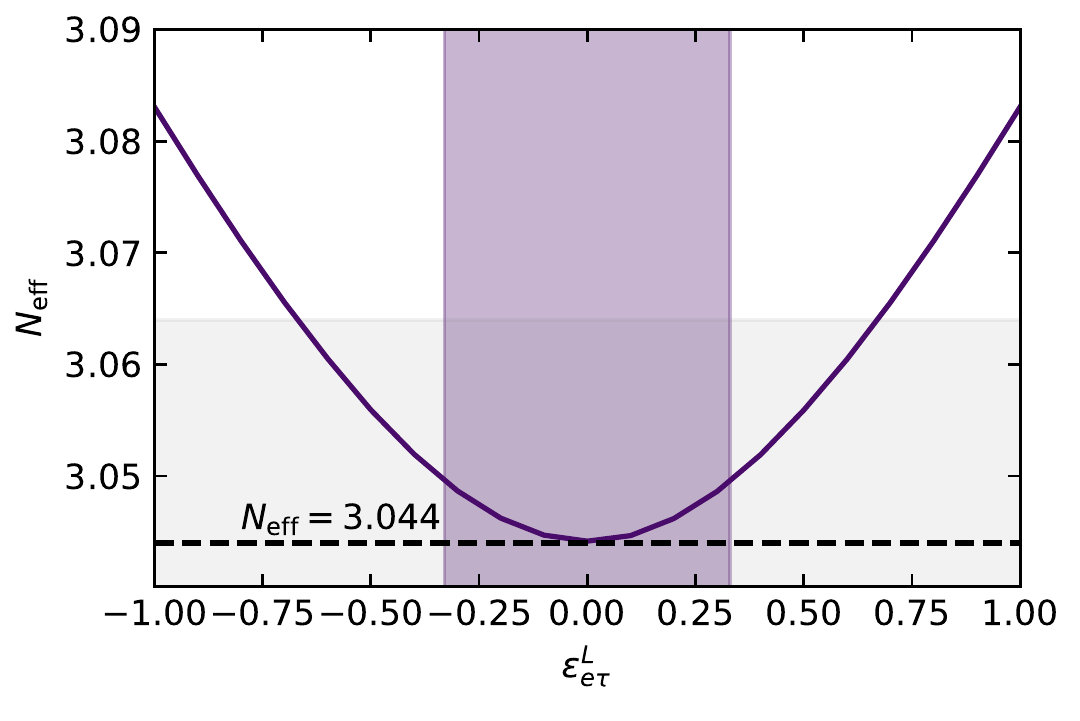}
    \caption{Values of $N_{\rm eff}$ in the presence of non-standard interactions. The left panel corresponds to non-universal NSI, with the effect of a non-zero $\varepsilon^L_{ee}$ in purple and that of a non-zero $\varepsilon^L_{\tau\tau}$ in pink. The right panel corresponds to flavour changing NSI, in particular, a non-zero $\varepsilon^L_{e\tau}$. The value of $N_{\rm eff} = 3.044$ expected in the absence of NSI is indicated by the black dashed line. The horizontal grey-shaded region corresponds to $\pm 0.02$ which is the expected sensitivity from future cosmological observations. The vertical coloured bands are the current terrestrial limits at 90$\%$ C.L. Both panels are adapted from Figures in \cite{deSalas:2021aeh} and reproduced with the authorisation of the original authors.}
    \label{fig:oneatatime}
\end{figure}

We first analyse the effect on non-universal NSI. In the left panel of Figure \ref{fig:oneatatime}, we show how $N_{\rm eff}$ varies in the presence of non-zero $\varepsilon^L_{ee}$ and $\varepsilon^R_{ee}$. One can see that for negative values of $\varepsilon^L_{ee}$, $N_{\rm eff}$ is found to be smaller than in the Standard Model. A minimum is found for $\varepsilon^ L_{ee} = -1/2 -\sin^2\theta_W$, when the shifted couplings in equations \eqref{eqn:shifted1} and \eqref{eqn:shifted3}  have a minimum. In the case of $\varepsilon^ L_{\tau\tau}$, the minimum is found, as expected, for $\varepsilon^L_{\tau\tau} = 1/2 -\sin^2\theta_W$, following the same argument.

Regarding flavour-changing NSI, one can see their impact on $N_{\rm eff}$ for the particular case of $\varepsilon^L_{e\tau}$ in the right panel of Figure \ref{fig:oneatatime}. Such NSI lead to higher value of $N_{\rm eff}$ due to the fact that they increase the strength of the interactions between neutrinos and electrons, as it can be seen in equations \eqref{eqn:shifted1} to \eqref{eqn:shifted3}. Moreover, the sign of the NSI parameter is expected to be irrelevant so the values of $N_{\rm eff }$ are expected to be symmetric with respect to $\varepsilon^ X_{\alpha\beta} = 0$. This is confirmed by the numerical calculations performed and shown in Figure \ref{fig:oneatatime}.

The numerical calculations performed and shown in Figure \ref{fig:oneatatime} validate the hypothesis that the main impact of NSI in the decoupling process is due to the changes in the scattering and in the annihilation between neutrinos and electrons. Besides that, in Ref.\ \cite{deSalas:2021aeh}, the contribution of NSI trough matter effects in oscillations to the value of $N_{\rm eff}$ was shown to be three orders of magnitude smaller than the contribution from the scattering and annihilation.

\begin{figure}
    \centering
    \includegraphics[width = 0.485\textwidth]{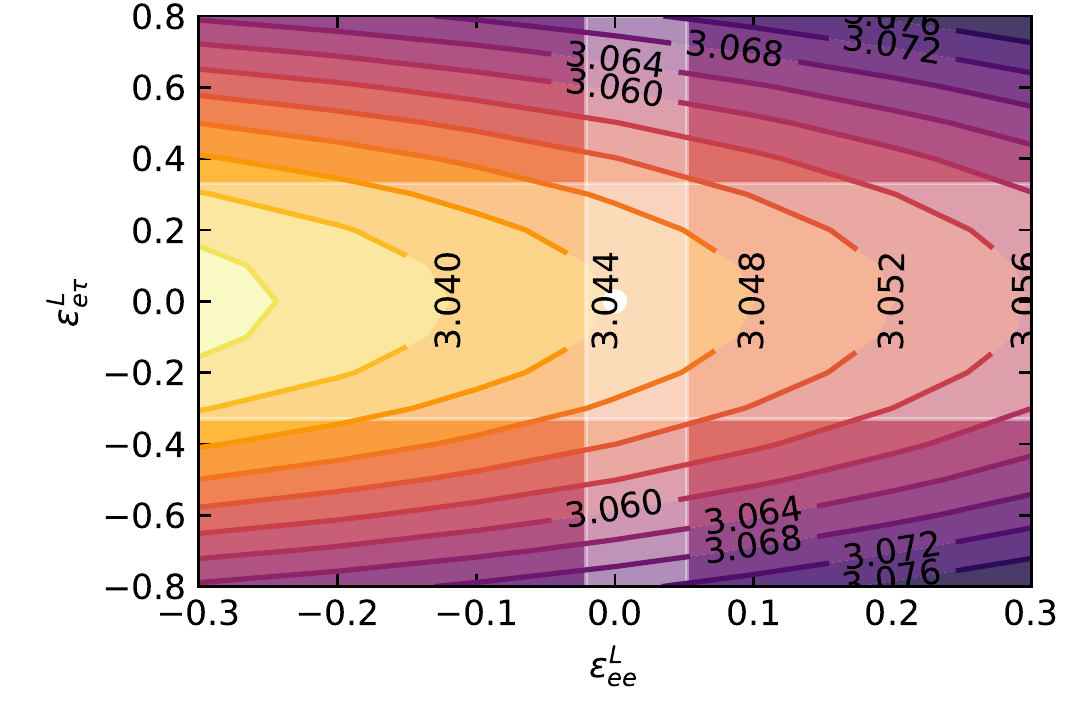}
    \includegraphics[width = 0.485\textwidth]{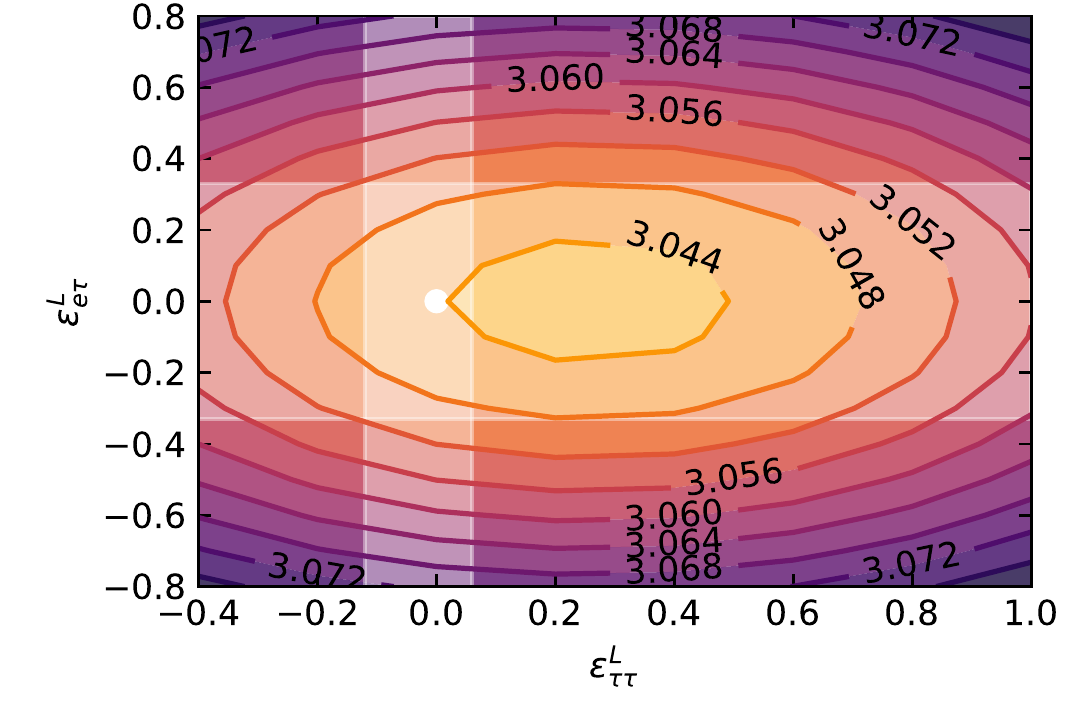}
    \caption{Isocontours of $N_{\rm eff}$ when two NSI parameters are allowed to vary simultaneously. The left panel shows the interplay between $\varepsilon^L_{ee}$ and $\varepsilon^L_{e\tau}$, and the right panel shows the combined effect of $\varepsilon^L_{\tau\tau}$ and $\varepsilon^L_{e\tau}$. The white point corresponds to the value of $N_{\rm eff} = 3.044$ in the absence of NSI. The white shaded bands are the 95$\%$ C.L. limits derived in terrestrial experiments varying one parameter at a time. Both panels are adapted from Figures in \cite{deSalas:2021aeh} and reproduced with the authorisation of the original authors. }
    \label{fig:twoatatime}
\end{figure}

It is also interesting to explore the possible interplay between pairs of NSI parameters. In Figure \ref{fig:twoatatime} we show the isocountours of $N_{\rm eff}$ when two NSI parameters are allowed to vary simultaneously. Notice that the contours of equal $N_{\rm eff}$ are ellipses in parameters space. Again, this can be well understood in the light of the shitfs introduced in the coupling by NSI parameters. Figure \ref{fig:twoatatime} also proves that an analysis varying two or more parameters at the same time is feasible, whereas many of the here presented terrestrial bounds only allowed for a non-zero NSI parameter at a time. 

Notice that although the expected constraints from future cosmological observations are not very competitive, they would provide an independent confirmation of the terrestrial constraints. In particular, it is in general complicated to constrain some sets of parameters simultaneously from terrestrial experiments, for instance $\varepsilon^X_{\tau\tau}$ and $\varepsilon^X_{e\tau}$, whereas we have shown that it is feasible using comsological data.

\section{Discussion}
In these proceedings, we have reported on a precision calculation of the process of neutrino decoupling in the presence of non-standard interactions between neutrinos and electrons. We found that the main impact of NSI is due to the change in the scattering and annihilation processes involving neutrinos and electrons, while matter effects altering neutrino oscillations play a very subleading role.

We have shown that, whereas flavour-changing NSI can only enhance the value of $N_{\rm eff}$, non-universal NSI can both increase or decrease the energy density transfer between neutrinos and electrons, which translates into a larger or smaller value of $N_{\rm eff}$ with respect to the value expected from the SM.

Finally, we commented on how the forecasted sensitivity of $\sigma(N_{\rm eff})= 0.02$ from future cosmological observatories will allow to constraint these non-standard interactions. The expected limits on NSI parameters would provide a complementary way to probe these interactions.

\ack
I would like to thank my collaborators Stefano Gariazzo, Pablo F.\ de Salas, Sergio Pastor and Mariam Tórtola for their contributions to the work here presented. I am also grateful for the hospitality of the Max-Planck-Institut f{\"u}r Kernphysik (Heidelberg) during the preparation of these proceedings.  This work is supported by FPU18/04571, FPA2017-85216P (AEI/FEDER, UE) and PROMETEO/2018/165 (Generalitat Valenciana). 

\section*{References}
\bibliographystyle{iopart-num}
\bibliography{bibliography}

\end{document}